\documentclass[%
 reprint,
 amsmath,amssymb,
 aps,
]{revtex4-2}
\usepackage{hyperref}
\usepackage{graphicx}
\usepackage{dcolumn}
\usepackage{bm}
\usepackage{orcidlink}

\begin{document}

\preprint{APS/123-QED}

\title{Search for lepton flavor-violating decay modes $B^0\to K_S^0\tau^\pm\ell^\mp$~$(\ell=\mu, e)$ with hadronic $B$-tagging at Belle and Belle~II}

  \author{I.~Adachi\,\orcidlink{0000-0003-2287-0173}} 
  \author{K.~Adamczyk\,\orcidlink{0000-0001-6208-0876}} 
  \author{L.~Aggarwal\,\orcidlink{0000-0002-0909-7537}} 
  \author{H.~Ahmed\,\orcidlink{0000-0003-3976-7498}} 
  \author{H.~Aihara\,\orcidlink{0000-0002-1907-5964}} 
  \author{N.~Akopov\,\orcidlink{0000-0002-4425-2096}} 
  \author{M.~Alhakami\,\orcidlink{0000-0002-2234-8628}} 
  \author{A.~Aloisio\,\orcidlink{0000-0002-3883-6693}} 
  \author{N.~Althubiti\,\orcidlink{0000-0003-1513-0409}} 
  \author{N.~Anh~Ky\,\orcidlink{0000-0003-0471-197X}} 
% ...
\collaboration{The Belle and Belle II Collaborations}

\begin{abstract}
We present the first search for the lepton flavor-violating decay modes $B^0 \rightarrow K_S^0 \tau^\pm \ell^\mp$~$(\ell=\mu, e)$ using the 711~fb$^{-1}$ and 365~fb$^{-1}$ data samples recorded by the Belle
and Belle~II detectors, respectively. We use a hadronic $B$-tagging technique, and search for the signal decay in the system recoiling against the fully reconstructed $B$ meson. We find no evidence for $B^0 \rightarrow K_S^0 \tau^\pm \ell^\mp$ decays and set 90\% confidence level upper limits on the branching fractions in the range of $[0.8,\,3.6]\times10^{-5}$.

\end{abstract}

\maketitle

Recent anomalies observed in semileptonic $B$ decays, particularly in transitions like $b \rightarrow c \tau \nu$~\cite{rd}, may indicate deviations from lepton flavor universality. According to this principle, the three generations of leptons are expected to interact identically with gauge bosons, except for mass differences. The experimental deviations from the standard model (SM) predictions suggest the potential existence of new heavy particles that couple preferentially to third-generation leptons. 
Recently, Belle~II reported a $b \to s\nu\bar \nu$ excess and obtained the branching fraction $\mathcal{B}(B^{+} \rightarrow K^{+} \nu \bar{\nu})=(2.3\pm0.7)\times 10^{-5}$~\cite{knn}, which is 2.7 standard deviations $(\sigma)$ larger than the SM expectation. If confirmed, this would not only be physics beyond the standard model (BSM), but could also reflect off-diagonal couplings between leptons of different flavors~\cite{lfvinb2s,b2sll}. Lepton flavor-violating (LFV) decays, which are forbidden in the SM, could then occur in $B$ meson decays at significant rates. Ref.\cite{b2sll}, starting from the $B^+ \to K^+ \nu \bar{\nu}$ excess, predicts an enhancement of $\mathcal{B}(B \to K \tau^\pm \mu^\mp)$ to $[2,\,3] \times 10^{-6}$, which is close to the current experimental sensitivity. The BSM models in Refs.~\cite{lfvinb2s} and~\cite{lfvinb} predict significant enhancements in the branching fractions $(\mathcal{B})$ of processes such as $b \to s\tau\ell$, with the effect being enhanced due to the coupling of the third-generation $b$ quark with the heaviest lepton, $\tau$. This results in a notable increase in the branching fractions of $B \to K \tau^\pm \ell^\mp~(\ell=\mu,e)$ decays and provides a potential experimental window into BSM physics. 

BaBar performed the first LFV searches in $B^{+} \rightarrow K^{+} \tau^{ \pm} \ell^{\mp}$ modes and set upper limits (ULs) on their branching fractions in the range of $[1.5,\,4.5] \times 10^{-5}$ at 90\% confidence level~(CL)~\cite{babar}. Belle provided the most stringent UL on the $B^{+} \rightarrow K^{+} \tau^{+} \mu^{-}$ decay of $6 \times 10^{-6}$ at 90\% CL using $772 \times 10^6$ $B\bar B$ pairs~\cite{ktauell}, significantly better than LHCb's limit of $3.9\times 10^{-5}$, obtained with 9~fb$^{-1}$ of $pp$ collision data~\cite{lhcb}. Moreover, LHCb set ULs on $B^0\to K^{* 0} \tau^{ \pm} \mu^{\mp}$ decays in the range of $[0.8,\,1.0]\times 10^{-5}$ at 90\% CL \cite{ksttauell}.

In this Letter, we present the first search for the LFV decays $B^0\to K_S^0\tau^\pm\ell^\mp~(\ell=\mu,e)$. We employ a hadronic $B$-tagging technique, and then use recoil mass to reconstruct the mass of the $\tau$. Our results are based on a combined analysis of $772 \times 10^6$ $B\bar{B}$ pairs (711 fb$^{-1}$) from Belle and $387 \times 10^6$ $B\bar{B}$ pairs (365 fb$^{-1}$) from Belle~II (2019--2022). The advantage of $K_S^0$ over $K^+$ and $K^{*0}$ in $B \to K\tau\ell$ decays is its very pure $K_S^0\to\pi^+\pi^-$ signature, which  is an additional advantage over $pp$ collision experiments, highlighting the unique strengths of Belle and Belle~II.

Belle operated at the KEKB asymmetric-energy collider with electron-(positron-)beam energies of 8.0(3.5)~GeV~\cite{kekb}. Belle~II operates at the successor, SuperKEKB, designed to deliver forty times higher instantaneous luminosity than KEKB, with electron-(positron-)beam energies of 7(4)~GeV~\cite{superkekb}. The Belle~II detector~\cite{b2detector} is a upgraded version of the Belle~\cite{bdetector}, including a vertex detector~(VXD), composed of two inner layers of pixel detectors~(PXD) and four outer layers of double-sided strip detectors~(SVD), a central drift chamber~(CDC), a time-of-propagation~(TOP) detector in the central detector volume and an aerogel ring-imaging Cherenkov~(ARICH) detector in the forward region, and an electromagnetic calorimeter (ECL). All these sub-detectors are located inside the same solenoid as in Belle, with a $K^0_L$–Muon detector (KLM) instrumented in the yoke.

The analysis procedure is first developed using simulation before being applied to data. EvtGen~\cite{EvtGen} is used to generate $e^{+} e^{-} \rightarrow \Upsilon(4S)\rightarrow B \bar{B}$ with final-state radiation simulated by PHOTOS~\cite{photons}. The $B^0\to K_S^0\tau^\pm\ell^\mp$ signal channels are modeled using an uniform three-body phase space model. The KKMC~\cite{kkmc} and PYTHIA~\cite{pythia} packages are used to simulate the $e^{+} e^{-} \rightarrow q \bar{q}$ continuum $(q=u, d, s, c)$. The detector responses are modeled by GEANT3~\cite{geant3} for Belle and GEANT4~\cite{geant4} for Belle~II. We use the Belle~II analysis software framework (basf2)~\cite{basf2} to reconstruct the events for both Belle and Belle~II data. The Belle data is converted to the Belle~II format for basf2 compatibility using the B2BII framework~\cite{b2bii}. 

In each $B\bar{B}$ pair, if one $B$ meson, $B_{\rm sig}$, decays to a final state involving neutrinos, it cannot be fully reconstructed as neutrinos escape the detector. However, the presence of missing energy and momentum can be inferred from the other $B$ meson. This process is called tagging, with the other $B$ meson referred to as $B_{\text{tag}}$. By combining the visible particles from $B_{\rm sig}$, we can kinematically constrain the undetected neutrinos as possible products of a $\tau$ decay. 
We require a fully reconstructed $B_{\rm tag}$ using the full-event-interpretation (FEI) algorithm \cite{fei}, a machine-learning based algorithm developed for $B$-tagging analyses at Belle and Belle~II. Each reconstructed $B_{\text{tag}}$ candidate is assigned a multivariate classifier output, $\mathcal{P}_{\mathrm{FEI}}$, ranging from zero (background-like) to one (signal-like). 
To constrain the $B_{\text{tag}}$ kinematics, we require the beam-energy-constrained mass $M_{\mathrm{bc}} = \sqrt{\left(E_{\text{beam}} / c^2\right)^2 - \left(p_{B_{\text{tag}}} / c\right)^2} > 5.27$~GeV/$c^2$, and the energy difference $\Delta E = E_{B_{\text{tag}}} - E_{\text{beam}}$ to satisfy $-0.15 < \Delta E < 0.1$~GeV. Here, $E_{\text{beam}}$, $E_{B_{\text{tag}}}$, and $p_{B_{\text{tag}}}$ denote the beam energy, and the energy and momentum of the $B_{\text{tag}}$ candidate in the $e^+e^-$ center-of-mass (c.m.) frame, respectively. If multiple $B_{\rm tag}$ candidates are reconstructed in a single event, the one with the highest \(\mathcal{P}_{\mathrm{FEI}}\) is chosen, and candidates satisfying \(\mathcal{P}_{\mathrm{FEI}} > 0.001\) are retained. Using these criteria, the $B$-tagging has an average efficiency of 0.59\% and a purity of 44\%. Here, purity is the ratio of reconstructed or expected signal events to total events in the signal region.

Tracks and clusters not associated to $B_{\rm tag}$ are used to reconstruct the $B_{\rm sig}$, whose flavor is assumed to be opposite to that of the $B_{\rm tag}$ candidate. The BSM couplings between $b\tau$ and $b\ell$, or between $s\tau$ and $s\ell$, can be different, leading to an asymmetric differential decay rate between $b \to s \tau^+ \ell^-$ and $b \to s \tau^- \ell^+$ (see Eq.~9 in Ref.~\cite{lfvinb2s}). To address this and the different background characteristics, we distinguish signal channels by the primary lepton charge and $b$ quark flavor: same-sign $SS_\ell$ ($B^0 \to K_S^0 \tau^- \ell^+$) and opposite-sign $OS_\ell$ ($B^0 \to K_S^0 \tau^+ \ell^-$). Due to $B^0$-$\bar{B}^0$ mixing, a small fraction ($\chi_d$)~\cite{pdg} of $SS_\ell$ decays are classified as $OS_\ell$ decays, and vice versa. The $\tau$ candidates are reconstructed via $\tau \rightarrow e \nu \bar{\nu}, \mu \nu \bar{\nu}, \pi \nu$, or $\rho(\rightarrow \pi\pi^0) \nu$, covering over 70\% of $\tau$ decays~\cite{pdg}. Signal channels are formed by combining a $K_S^0$, a primary lepton ($\ell$), and a $\tau$ decay track ($t_\tau$). 

We reconstruct $K_S^0$ candidates from a pair of oppositely charged tracks assumed to be pions, with a common vertex. We use a standard momentum-binned $K_S^0$ selection that includes requirements on the $K_S^0$ flight information~\cite{goodbellekshort}. The purity of the $K_S^0$ candidates exceeds 98\%, with backgrounds predominantly containing real \(K_S^0\) candidates. 

To select $\ell$ and $t_\tau$, we require the transverse ($d_0$) and longitudinal ($z_0$) projection of the distance of closest approach to the origin to be less than $0.5 \mathrm{~cm}$ and $5.0 \mathrm{~cm}$ to reduce misreconstructed or spurious tracks from beam-induced background. At least 20 hits in the CDC are required. For Belle, we use information from the KLM only to identify muon candidates, while for Belle II, we use information from all sub-detectors except the VXD. Muon candidates are required to have momenta greater than $0.6~\mathrm{GeV}/c$ to sufficiently penetrate the KLM. This selection has an efficiency of 89\% with a pion misidentification rate lower than 2.5\% for both samples~\cite{muid_b1}. Electrons are required to have momenta greater than 0.3~GeV/$c$ to lie in the acceptance of the ECL. For Belle, electrons are identified using the information from the ECL, CDC and aerogel threshold Cerenkov counter (ACC), respectively. For Belle~II, the electron identification uses a boosted-decision-tree (BDT) classifier trained with information from all sub-detectors except the VXD. The electron identification has an efficiency of 92\% (86\%) and a pion misidentification rate below 0.3\% (0.5\%) for Belle (Belle~II)~\cite{eid_b1}. To recover electron candidates with bremsstrahlung, we accept photons with a minimum energy of 50 MeV that are within a 50 mrad angle of an electron track. Pion candidates for $t_\tau$ reconstruction are selected using PID likelihoods using information from the ACC, CDC, and time-of-flight scintillation counters for Belle. For Belle~II, information from all the subdetectors except the VXD is used. This achieves a pion identification efficiency of 85\% (83\%) and a kaon misidentification rate of 6\% (8\%) for Belle (Belle~II). 

The $\tau \rightarrow \rho(\rightarrow \pi \pi^0) \nu$ mode, the most probable $\tau$ decay channel ($\mathcal{B}>20\%$), has never been used in $B \rightarrow K \tau \ell$ analyses \cite{babar, lhcb, ktauell}. Understanding this mode is crucial for improving the kinematic properties of $t_\tau$ and enhancing background rejection strategies. However, its reconstruction is challenging due to contamination from \(\pi^0\) mesons, which can be misreconstructed using either a fake photon (clusters associated with hadronic deposits) or a photon from beam-background. To reconstruct the clean \(\tau \rightarrow \rho\nu\) mode, photons are selected with energies above 50(60)~MeV, 100(75)~MeV, and 150(100)~MeV in the barrel, forward, and backward endcaps for Belle (Belle~II). 
We developed classifiers to suppress these backgrounds using BDTs. The common cluster features are energy, polar angle (relative to the beam-pipe), lateral energy distribution~\cite{lat}, distance between the cluster and its nearest track, and fraction of cluster energy detected in the central part. For Belle~II classifiers, we include additional features: the time-difference between the collision and reconstructed cluster; outputs of classifiers using eleven Zernike moments~\cite{zernike}; and identifiers for electromagnetic or hadronic showers using pulse shape discrimination~\cite{pulseshape}. For Belle, additional features include azimuthal angle, number of crystals in the cluster, and energy in the most energetic crystal. On average, the classifiers reduce backgrounds by 90\% with a $\pi^0$ efficiency of 70\%, including the selection $0.125 < M_{\pi^0} < 0.145~{\rm GeV}/c^2$. The $\rho$ candidates are selected with $0.60 < M_{\pi\pi^0} < 0.94~{\rm GeV}/c^2$, with one candidate randomly chosen due to the multiplicity of 1.07.

The rest-of-event (ROE) consists of the tracks and clusters not used in $B_{\rm tag}$ and $B_{\rm sig}$. We select events without any track in the ROE having $|d_0|<10$~cm and $|z_0|<20$~cm. The ROE clusters are required to satisfy the same selection criteria as the photons used for $\pi^0$ reconstruction. If the $t_{\tau}$ candidate is identified with multiple particle hypotheses, we assign a single one according to the following priority order (based on the purity of the modes): muon, electron, and pion. We require the $\tau \to \pi \nu$ mode to have no additional $\pi^0$ candidate in the ROE to avoid double counting with the $\tau \to \rho \nu$ mode.

The $B_{\rm sig}$ momentum is equal in magnitude and opposite in direction to that of $B_{\rm tag}$, $\vec{p}_{B_{\rm tag}}$, and the $B_{\rm sig}$ energy is equal to $E_{\rm beam}$ in the c.m. frame. Therefore, the $\tau$ momentum and energy are given by,
\begin{align*}
     \vec{p}_\tau & =-\vec{p}_{B_{\mathrm{tag}}}- \vec{p}_{K_S^0}- \vec{p}_{\ell}, \\
E_\tau & =E_{\rm {beam}}-E_{K_S^0}-E_{\ell},
\end{align*}
from which we reconstruct the recoiling $M_{\tau}$ according to Eq.~\ref{eq:mrecoiling}. The signal yields are then extracted from $M_\tau$, as signal events peak at the known $\tau$ mass~\cite{pdg}, while the background remains flat without any peaking structures in the simulation. When $t_\tau$ is $\mu$ or $e$ (i.e.\ when there are two leptons in $B_{\rm sig}$), it is possible to form both $SS_\ell$ and $OS_\ell$ candidates, but this does not bias the signal yield as misassigned candidates do not peak in $M_{\tau}$.

\begin{equation}\label{eq:mrecoiling}
\begin{split}
M_{\rm recoil} = M_{\tau} = \Big[ m_{B}^{2} + M_{K_S^0 \ell}^{2} - 2 \Big( E_{\rm beam} E_{K_S^0 \ell} \\
+ |\vec{p}_{B_{\rm tag}}| |\vec{p}_{K_S^0 \ell}| \cos \theta \Big) \Big]^{\frac{1}{2}}
\end{split}
\end{equation}
Here, $m_{B}$ is the known $B^0$ mass~\cite{pdg}; $M_{K_S^0 \ell}$, $E_{K_S^0 \ell}$, $\vec{p}_{K_S^0 \ell}$ are the mass, energy, and momentum of the system composed of the $K_S^0$ and $\ell$, respectively; $\theta$ is the angle between $\vec{p}_{{B}_{\rm tag}}$ and $\vec{p}_{K_S^0 \ell}$. The $M_{\tau}$ resolution is approximately 25~MeV/$c^2$ for both Belle and Belle~II simulations. 

The background after the pre-selection is mainly decays arising from $b\to c$ transitions. Events with $B^0\to D^{(*)-}(\to K_S^0 t^- X)\ell^+\nu$ decays, where the reconstructed primary lepton originates from semileptonic $B^0$ decay and a track $t^-$ from $D^{(*)-}$ is misinterpreted as coming from the $\tau$ decay,  with $X$ representing any other particles, constitute the dominant source of background in $SS_\ell$ modes. If $B^0 \to D^{(*)-}\ell^+\nu$ undergoes a flavor transformation due to $B^0$-$\bar{B}^0$ mixing, this results in a change in the signs of the final states ($\bar{B}^0\to D^{(*)+}\ell^-\bar{\nu}$) and provides the appropriate sign configuration for the final state in $OS_\ell$ modes. Additionally, the reconstruction of the primary lepton from $B_{\rm sig}$ tends to favor a higher momentum lepton originating from the $B^0$ meson. Consequently, semileptonic $B$ decays are the primary background in both $SS_\ell$ and $OS_\ell$ modes. As the $K_{S}^0$ and $t_{\tau}$ come from a $D$ meson, we require the invariant mass of $K^0_{S}$ and $t_{\tau}$ to be greater than 1.91~GeV/$c^2$, i.e., greater than the $D$ meson mass accounting for resolution. For $\tau\to\rho\nu$ decay, we require $M_{K_S^0\rho}$ to be greater than 2.1~GeV/$c^2$ due to the poor mass resolution resulting from the presence of a $\pi^0$ candidate. Because $B^0\to K_S^0 J/\psi (\to\ell^+\ell^-)$ background can pass our selection criteria, events in the range $3.00<M_{t_\tau\ell}<3.14~{\rm GeV}/c^2$ are rejected when $\ell$ is an electron (muon) for the $\tau\to e (\mu)\nu\bar\nu$ mode. To suppress the photon conversion background, we require $M_{t_\tau\ell}$, in this case $M_{e^+e^-}$, to be greater than $0.15~{\rm GeV}/c^2$.

Continuum $q\bar q$ events can be distinguished from $B\bar B$ events by exploiting their difference in event topologies. We use sphericity-related variables~\cite{sphericity} and require the cosine of the angle between the thrust axis of $B_{\rm tag}$ and the other particles not used in $B_{\rm tag}$ ($\rm cos~\theta_{\rm T}$) to be less than 0.9. These selections reduce the $q\bar q$ background by 86\% and retain 88\% of the signal for both samples. 

After the above selection criteria, the background consists of charm meson semileptonic decays with a $K_S^0$ and $\ell$ in their final state, other $B\bar B$ decays, and $q\bar q$ events. For each signal mode, BDT classifiers are trained with 11 features to suppress the residual backgrounds. These features include $M_{K_S^0\ell}$, which helps to suppress the background from charm meson semileptonic decays, the sum of ECL cluster energies in the ROE, energies of the $\ell$ and $t_\tau$, event-shape variables and modified Fox-Wolfram moments~\cite{KSFW}, which help to reduce the $q\bar q$ background. The same BDT output criterion are used for Belle and Belle~II because of the similar performance, determined using figure of merit ${\epsilon_{\rm sig}}/{(\frac{a}{2}+\sqrt{N_{\rm bkg}}})$~\cite{punzi}. Here, $\epsilon_{\rm sig}$ is the signal efficiency, $N_{\rm bkg}$ is the expected background yield and $a=3$ represents the target significance in terms of standard deviations. The BDT selection results in an average signal efficiency of about 75\% and rejects 90\% of the remaining background.

The purity in the simulations is similar in Belle and Belle~II for every channel after applying all selection criteria, so the two datasets are merged. To extract the signal yield, we do a single unbinned-maximum-likelihood fit to the \(M_\tau\) distribution of the combined dataset. The probability density function (PDF) used to model the $M_\tau$ signal distribution is a Johnson function~\cite{johnson}. The parameters that describe the signal shape are fixed to the values obtained from the fit to the simulated samples. Background events have a smooth distribution in the $M_{\tau}$ fit region, modeled using a second-order polynomial. 
To validate the fitting procedure, we generate large ensembles of simulated experiments, in which the $M_\tau$ distributions are produced from the PDFs used for fitting. Comparisons of the simulated and measured signal yields indicate no obvious bias.

The $B^0\to D^-\pi^+$ sample is used to calibrate $B_{\rm tag}$ efficiency. We reconstruct $B_{\rm tag}$ using the FEI and a high momentum $\pi^+$, then compute $M_{\rm recoil}$ to observe the $D$ signal using Eq.~\ref{eq:mrecoiling}. The yield ratio $\mathcal{R}_{\rm FEI}=N_{\rm data}/N_{\rm simulation}=0.74\pm0.04~(0.81\pm0.04)$ is taken as the calibration factor for the $B_{\rm tag}$ efficiency in Belle (Belle~II). 

\begin{figure}[htp]
    \centering
\includegraphics[width=0.75\linewidth]{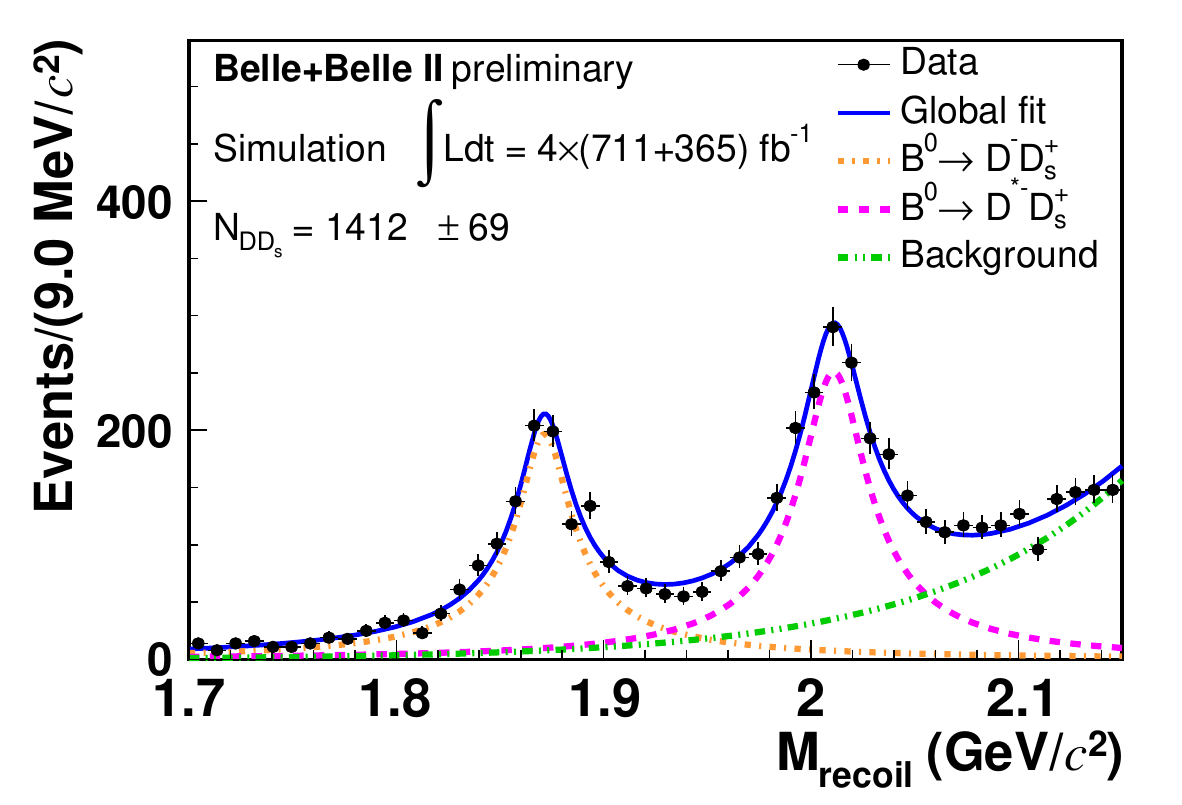} \\
\includegraphics[width=0.75\linewidth]{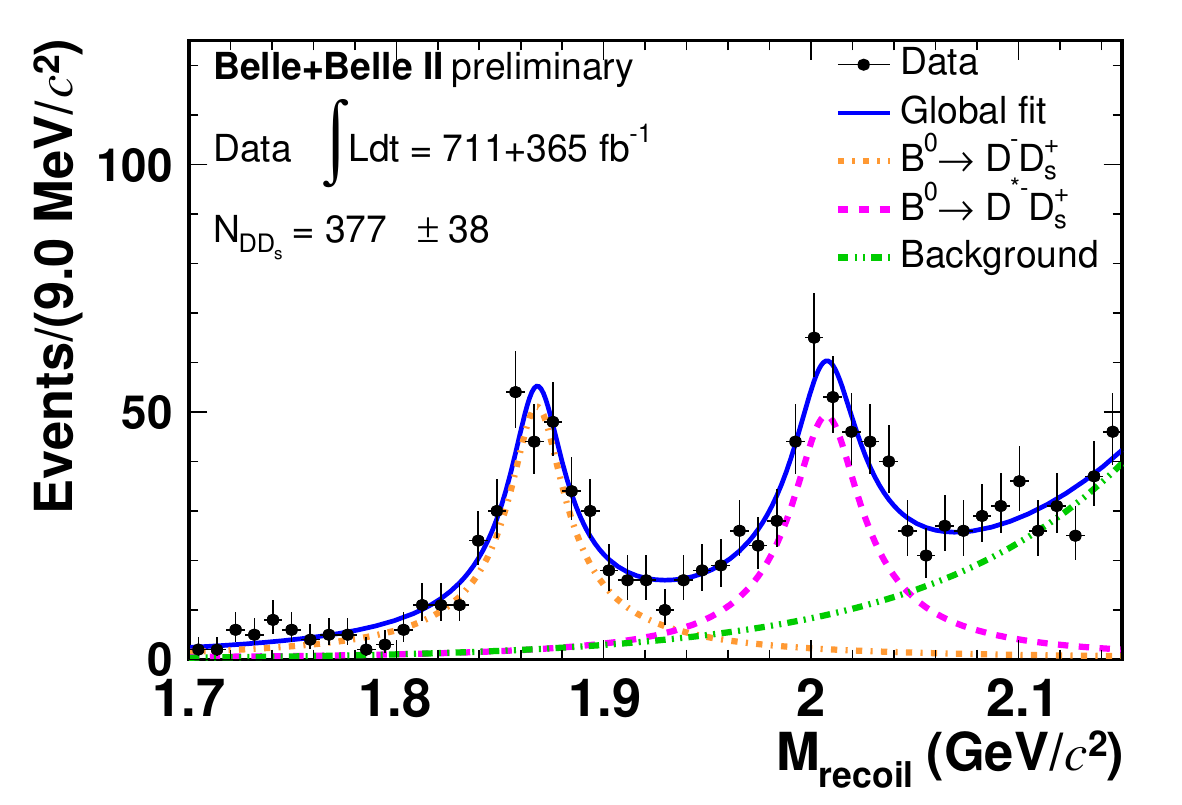}
\caption{Fit to the recoil mass of the $B_{\mathrm{tag}}$ and $D_s^{+}$ system for simulation (upper) and data (lower) with the combined Belle and Belle~II samples.}
    \label{fig:signalpdf}
\end{figure}

The $B^0 \to D_s^+ D^-$ sample is used to calibrate the signal PDF and BDT selections. The $D^-$ mass is reconstructed similarly to the $\tau$ mass in the signal decays. We reconstruct neutral $B_{\rm tag}$ candidates using the FEI algorithm and $D_s^+$, sharing the same momentum range as $K_S^0\ell$, through the decays $\phi(\to K^+K^-)\pi^+$ and $K_S^0 K^+$ and a charged track from $D$ as $t_\tau$. The distribution of the mass recoiling against the $B_{\rm tag}$ and $D_s^+$ system is shown in Fig.~\ref{fig:signalpdf} and clear signals for $D^{-}$ and $D^{*-}$ are visible. The $B^0\to D_s^+D^{*-}$ component is also fitted but we do not use it due to its lower purity. The $B^0\to D_s^+D^-$ signal PDF is modeled by the Johnson function and the parameters are fixed in the fits, except the mean, while the background is described by an exponential function with floating yield and shape parameters. We introduce a scale factor $f$ to account for data-simulation difference on the width from signal simulation. The $f$ ratio, determined to be $1.04 \pm 0.15$, is used as a correction factor for the width. The value of $\mathcal{B}(B^0\to D_s^+ D^-)$ is measured to be $(10.1\pm 1.2)\times 10^{-3}$, consistent with the world-average value~\cite{pdg} within 2$\sigma$, and serves as a closure test of the entire analysis chain. To validate the BDT performance, we apply $B^0\to K_S^0\tau^\pm\ell^\mp$ weights to $B^0\to D_s^+D^-$ events. The efficiency is derived from $B^0\to D_s^+D^-$ yields before and after BDT selection using $M_{\rm recoil}$ fits. The data-simulation efficiency ratios ($\mathcal{R}_{\rm BDT}$) are $0.93\pm0.17$, $0.96\pm0.16$, $0.92\pm0.16$, and $0.96\pm0.18$ for the $OS_\mu$, $SS_\mu$, $OS_e$, and $SS_e$ modes, respectively.

\begin{figure}[h]
\centering
\includegraphics[width=0.494\linewidth]{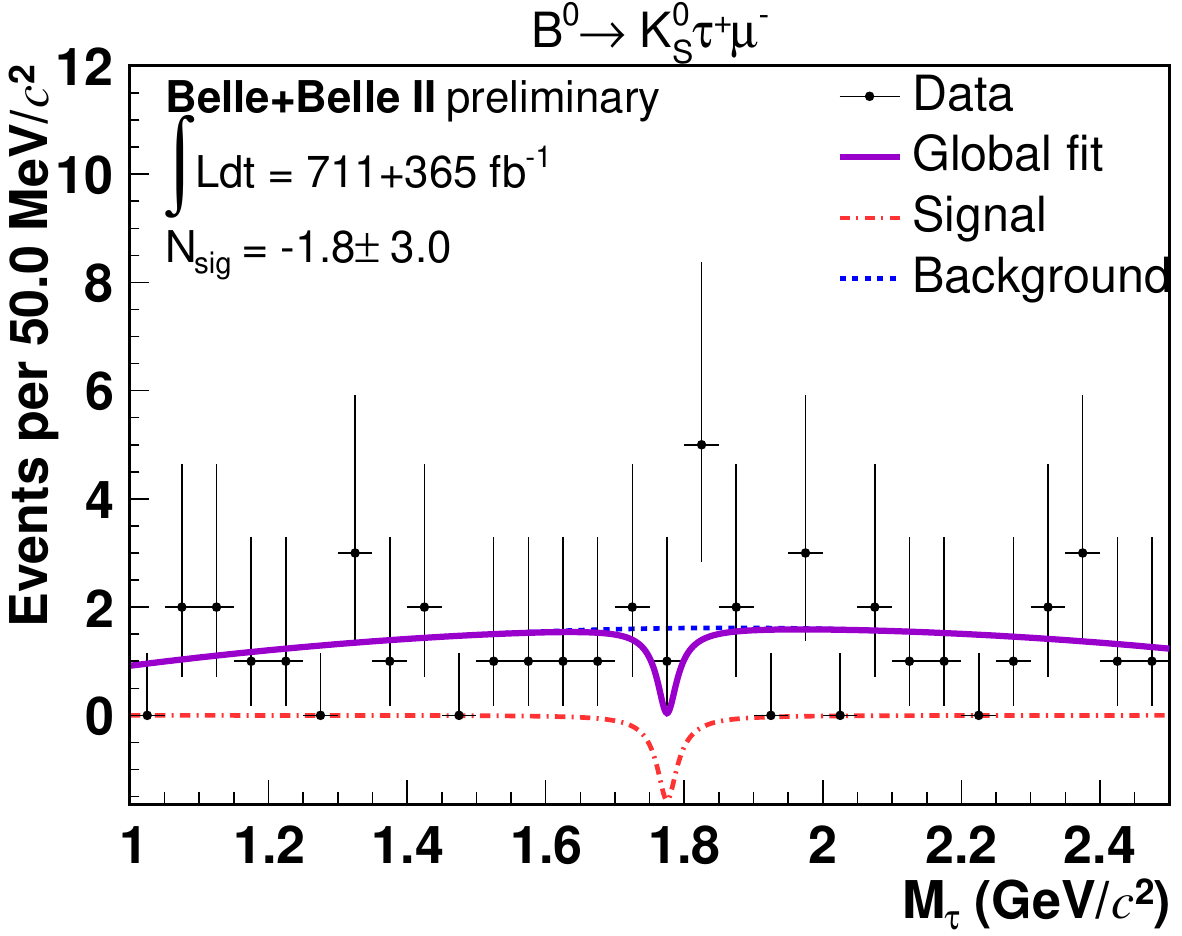}
\includegraphics[width=0.494\linewidth]{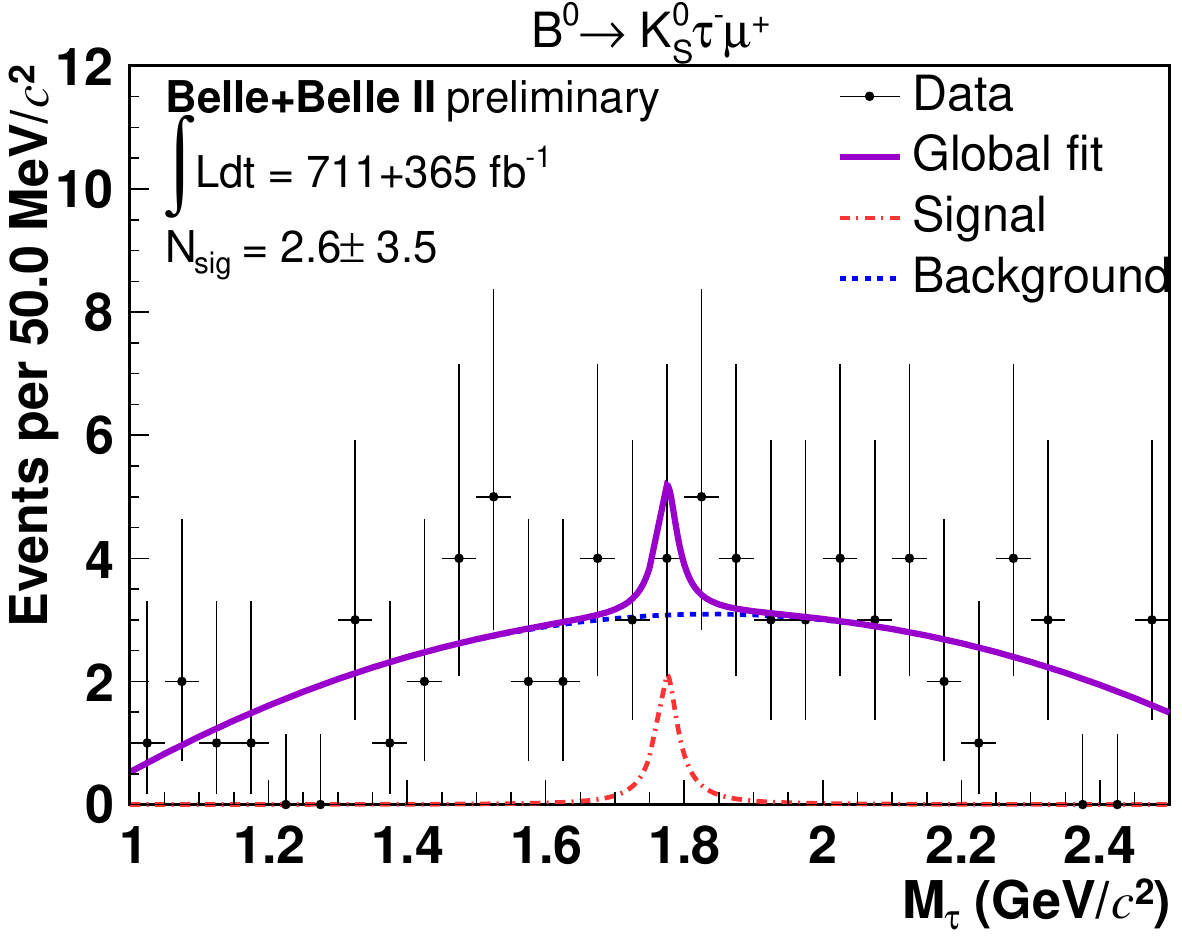}
\includegraphics[width=0.494\linewidth]{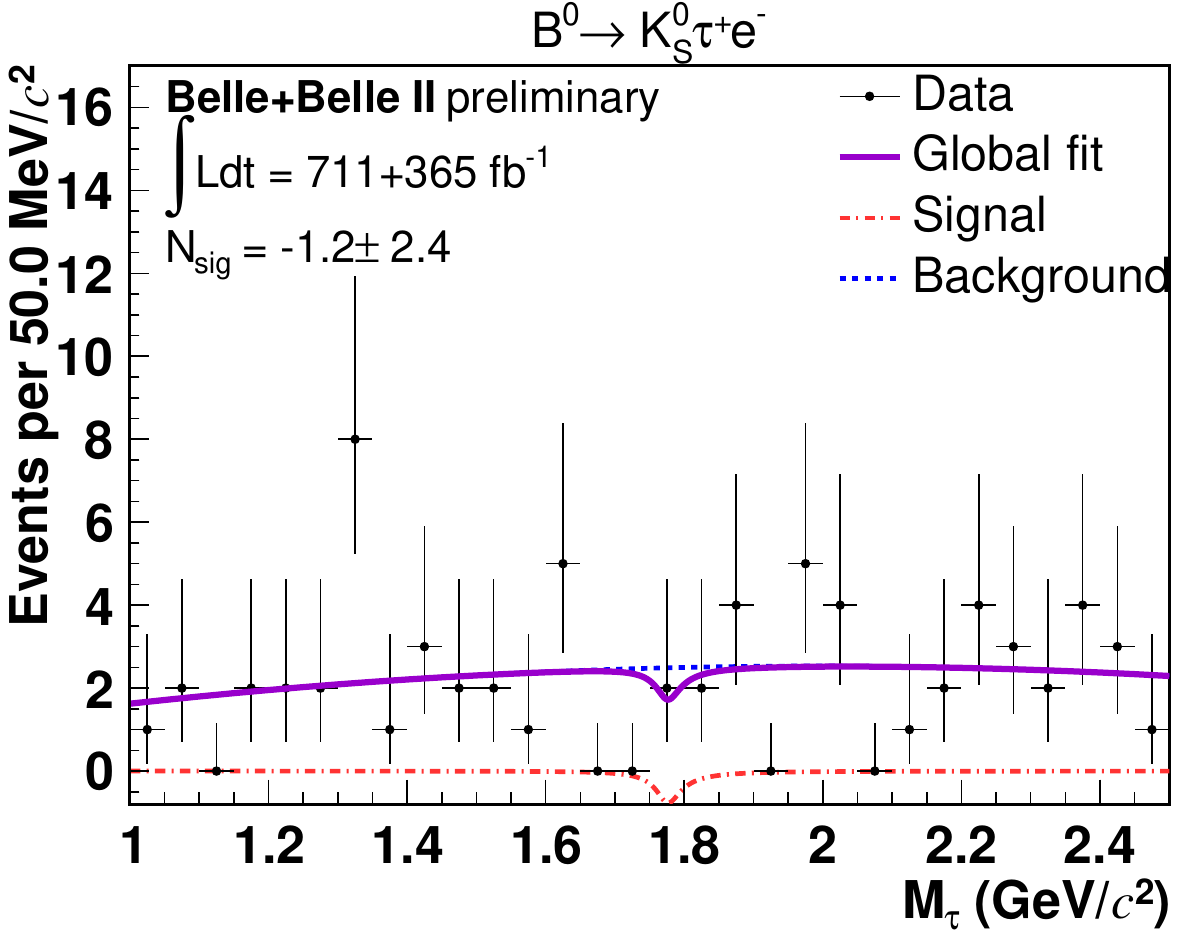}
\includegraphics[width=0.494\linewidth]{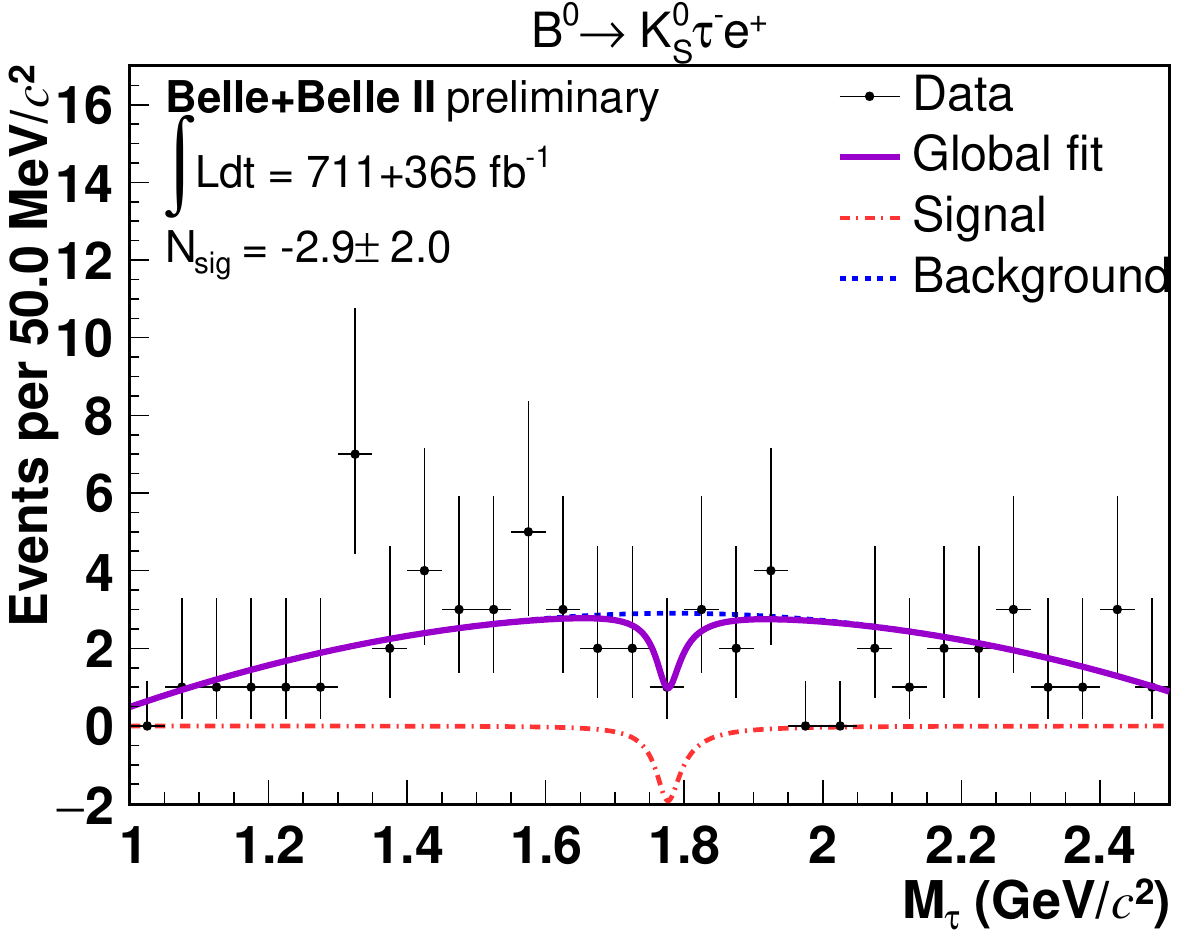}
\caption{
The $M_\tau$ distributions and fits for the combined Belle and Belle~II datasets. The black dots with error bars show the data, the red dash-dotted curve shows the signal component, the blue dashed curve shows the background component, and the purple solid curve shows the global fit.
}
\label{fig:finalfit}
\end{figure}

Figure~\ref{fig:finalfit} shows the $M_\tau$ fits to data for $B^0 \to K^0_S\tau^\pm\ell^\mp$ decays. There is no significant signal in any of the fit channels. We translate the number of observed events $N_{\rm{sig}}$ into a branching fraction $\mathcal{B}$ using the expression 
\begin{equation}
\label{eq:ul}
\mathcal{B} \\
=\frac{N_{\rm s i g}}{\epsilon \times 2 N_{B \bar{B}}\times(1+f_{+-}/f_{00})^{-1}},
\end{equation}
where $\epsilon$ is the efficiency after $\mathcal{R}_{\rm BDT}$ and $\mathcal{R}_{\rm FEI}$ calibrations. The efficiency also includes the branching fractions of \(K_S^0\), \(\tau\), \(\rho\), \(\pi^0\), and the effect of \(B^0\)-\(\bar B^0\) mixing (i.e. signal loss in mixed events) in the simulation. In the case where the true branching fractions are zero, the resulting estimates are unbiased. We use $N_{B\bar B}$= $1159 \times 10^{6}$, which is the total number of $BB$ pairs for the combined datasets; and $f_{+-}/f_{00} = 1.052\pm 0.031$, which is the ratio of $\mathcal{B}(\Upsilon(4S)\to B^+ B^-)$ to $\mathcal{B}(\Upsilon(4S)\to B^0\bar{B}^0)$~\cite{f+-00}. 

We obtain ULs on the signal yields using pseudo-experiments. They are generated using background and signal PDFs for different values of the signal branching fractions, performing 10,000 fits for each value. We then define $N_{\rm sig}^{\rm UL}$ at 90\% CL as the signal yield for which 10\% of the experiments have fit yields less than the observed $N_{\rm sig}$ in data. Systematic uncertainties are included by smearing the $N_{\rm{sig}}$ distribution obtained from the pseudo-experiments with the fractional systematic uncertainty, which has an effect of less than 1\% on the mean $N_{\rm{sig}}$. The ULs on the branching fractions $\mathcal{B}^{\text{UL}}$ are then obtained from $N_{\text{sig}}^{\text{UL}}$ using Eq.~\ref{eq:ul}. Including the effect of $B^0$-$\bar B^0$ mixing in the efficiency (Eq.~\ref{eq:ul}) ensures ULs correctly cover the case of zero true branching fractions and are conservative otherwise. Table~\ref{tab:results} summarizes the efficiency, fit results, and observed ULs at 90\% CL for the four channels. Expected ULs, derived from the no-signal assumption, are in the range $[2.1,2.2]\times 10^{-5}$. 

\begin{table}[h]
  \renewcommand{\arraystretch}{1.3}
    \caption{Efficiencies ($\epsilon$), signal yields ($N_{\rm sig}$) of the data fit, central value of the branching fractions and the observed $\mathcal{B}^{\rm UL}$ at 90\% CL. The first uncertainty of the central value is statistical and the second is systematic.}
    \centering
      \setlength{\tabcolsep}{2.5pt} 
    \begin{tabular}{cccccc}\hline\hline
        && &  \multicolumn{2}{c}{$\mathcal{B}(10^{-5})$} \\
  Channels & $\epsilon(10^{-4})$ & $N_{\rm sig}$ &  Central value & UL\\\hline
    $B^0 \rightarrow K_S^0 \tau^{+} \mu^{-}$& 1.7 & $-1.8\pm{3.0}$ & $-1.0\pm {1.6}\pm0.2$& 1.1\\ 
    $B^0 \rightarrow K_S^0 \tau^{-} \mu^{+}$& 2.1 & $~~2.6\pm3.5$ &~~$1.1\pm{1.6}\pm0.3$& 3.6\\ 
    $B^0 \rightarrow K_S^0 \tau^{+} e^{-}$& 2.0 & $-1.2\pm2.4$ &$-0.5\pm1.1\pm0.1$& 1.5\\ 
    $B^0 \rightarrow K_S^0 \tau^{-} e^{+}$& 2.1 & $-2.9\pm2.0$ &$-1.2\pm0.9\pm0.3$& 0.8\\\hline\hline 
    \end{tabular}
    \label{tab:results}
\end{table}

The primary source of systematic uncertainty arises from the BDT selections, which is 16--18\%, based on the uncertainty in $\mathcal{R}_{\rm BDT}$ using the $B^0\to D_s^+D^-$ sample. Using the same sample, the uncertainty from the signal PDF is 15\%. This includes the uncertainties in width (uncertainty of the width correction factor), mean (deviation from nominal $D$ mass in the data fit), skewness, and Gaussian component strength of the Johnson function, estimated using a new PDF reweighted by mode-dependent calibration factors for the dominant $B$-tagging modes. The uncertainty in the $B_{\rm tag}$ efficiency is taken from the uncertainty of $\mathcal{R}_{\rm FEI}$ (4\%). The small difference (0.8--1.6$\%$) in the validation of the fitting procedure is treated as the associated uncertainty. The uncertainty in $K_S^0$ reconstruction is estimated to be 1.1\% using a $D^{*+}\to\pi^+ D^0, D^0\to K_S^0\pi^+\pi^-$ sample. The Belle PID uncertainties are evaluated using $J / \psi \rightarrow \ell^{+} \ell^{-}$ and $D^{*+} \rightarrow$ $D^0 (\to K^{-} \pi^{+}) \pi^{+}$ samples to be 0.3\%, 0.4\% and 1.0\% for muons, electrons, and pions, respectively. The Belle~II PID uncertainties for muon, electron and pion are 0.5\%, 1.0\% and 1.0\%, respectively, which are obtained using the samples described in Ref.~\cite{sys}. The uncertainty from the $\pi^0$ reconstruction is 1.3\% using $B^+\to K^{*+}(\to K^+\pi^0)J/\psi$ and $D^{*-}\to {\bar D^0}(\to K^+\pi^-\pi^0)\pi^-$ samples. The uncertainty for the requirement that there is no additional $\pi^0$ candidate in the ROE in the $\tau\to\pi\nu$ mode is 1.0\% using $B_{\rm tag}B(\to K_S^0J/\psi)$ events. 
The uncertainties arising from $N_{B \bar{B}}$, $f_{+-}/f_{00}$, and the branching fractions of $K_S^0,\tau,\rho$ and $\pi^0$ decays~\cite{pdg} are $1.1 \%$, 1.5\% and 0.7\%, respectively. For sources with different systematic uncertainties in Belle and Belle~II, we calculate the total multiplicative values by weighting the individual uncertainties according to the integrated luminosities of the two samples. The total systematic uncertainties are 24\%, 22\%, 23\%, and 24\% for $OS_\mu$, $SS_\mu$, $OS_e$, and $SS_e$ modes, respectively.

In summary, we have searched for $B^0 \to K_S^0 \tau^\pm \ell^\mp$ for the first time using Belle and Belle~II datasets. This is also the first direct search for LFV in $B$ decays using the Belle~II dataset. 
The ULs on the branching fractions at 90\% CL are:
\begin{align*}
\mathcal{B}(B^0\to K_S^0\tau^+\mu^-) < 1.1\times10^{-5}\\
\mathcal{B}(B^0\to K_S^0\tau^-\mu^+) < 3.6\times10^{-5}\\
\mathcal{B}(B^0\to K_S^0\tau^+e^-) < 1.5\times10^{-5}\\
\mathcal{B}(B^0\to K_S^0\tau^-e^+) < 0.8\times10^{-5}
\end{align*}

The results for $B^0 \to K_S^0 \tau^\pm e^\mp$ are the most stringent ULs on $b \to s \tau e$ transitions, and those for $B^0 \to K_S^0 \tau^\pm \mu^\mp$ are among the best limits on $b \to s \tau \mu$ transitions achieved to date. These results are approaching the potential BSM enhancement level of $\mathcal{O}(10^{-6})$. Additionally, we provide the selection efficiency as a function of $(M_{\tau\ell}^2, M_{K^0_S \ell}^2)$ in the Supplemental Material, to allow these results to be reinterpreted in specific BSM models.

\begin{acknowledgments}

This work, based on data collected using the Belle II detector, which was built and commissioned prior to March 2019,
and data collected using the Belle detector, which was operated until June 2010,
was supported by
Higher Education and Science Committee of the Republic of Armenia Grant No.~23LCG-1C011;
Australian Research Council and Research Grants
No.~DP200101792, 
No.~DP210101900, 
No.~DP210102831, 
No.~DE220100462, 
No.~LE210100098, 
and
No.~LE230100085; 
Austrian Federal Ministry of Education, Science and Research,
Austrian Science Fund
No.~P~34529,
No.~J~4731,
No.~J~4625,
and
No.~M~3153,
and
Horizon 2020 ERC Starting Grant No.~947006 ``InterLeptons'';
Natural Sciences and Engineering Research Council of Canada, Compute Canada and CANARIE;
National Key R\&D Program of China under Contract No.~2022YFA1601903,
National Natural Science Foundation of China and Research Grants
No.~11575017,
No.~11761141009,
No.~11705209,
No.~11975076,
No.~12135005,
No.~12150004,
No.~12161141008,
No.~12475093,
and
No.~12175041,
and Shandong Provincial Natural Science Foundation Project~ZR2022JQ02;
the Czech Science Foundation Grant No.~22-18469S 
and
Charles University Grant Agency project No.~246122;
European Research Council, Seventh Framework PIEF-GA-2013-622527,
Horizon 2020 ERC-Advanced Grants No.~267104 and No.~884719,
Horizon 2020 ERC-Consolidator Grant No.~819127,
Horizon 2020 Marie Sklodowska-Curie Grant Agreement No.~700525 ``NIOBE''
and
No.~101026516,
and
Horizon 2020 Marie Sklodowska-Curie RISE project JENNIFER2 Grant Agreement No.~822070 (European grants);
L'Institut National de Physique Nucl\'{e}aire et de Physique des Particules (IN2P3) du CNRS
and
L'Agence Nationale de la Recherche (ANR) under grant ANR-21-CE31-0009 (France);
BMBF, DFG, HGF, MPG, and AvH Foundation (Germany);
Department of Atomic Energy under Project Identification No.~RTI 4002,
Department of Science and Technology,
and
UPES SEED funding programs
No.~UPES/R\&D-SEED-INFRA/17052023/01 and
No.~UPES/R\&D-SOE/20062022/06 (India);
Israel Science Foundation Grant No.~2476/17,
U.S.-Israel Binational Science Foundation Grant No.~2016113, and
Israel Ministry of Science Grant No.~3-16543;
Istituto Nazionale di Fisica Nucleare and the Research Grants BELLE2,
and
the ICSC – Centro Nazionale di Ricerca in High Performance Computing, Big Data and Quantum Computing, funded by European Union – NextGenerationEU;
Japan Society for the Promotion of Science, Grant-in-Aid for Scientific Research Grants
No.~16H03968,
No.~16H03993,
No.~16H06492,
No.~16K05323,
No.~17H01133,
No.~17H05405,
No.~18K03621,
No.~18H03710,
No.~18H05226,
No.~19H00682, 
No.~20H05850,
No.~20H05858,
No.~22H00144,
No.~22K14056,
No.~22K21347,
No.~23H05433,
No.~26220706,
and
No.~26400255,
and
the Ministry of Education, Culture, Sports, Science, and Technology (MEXT) of Japan;  
National Research Foundation (NRF) of Korea Grants
No.~2016R1-D1A1B-02012900,
No.~2018R1-A6A1A-06024970,
No.~2021R1-A6A1A-03043957,
No.~2021R1-F1A-1060423,
No.~2021R1-F1A-1064008,
No.~2022R1-A2C-1003993,
No.~2022R1-A2C-1092335,
No.~RS-2023-00208693,
No.~RS-2024-00354342
and
No.~RS-2022-00197659,
Radiation Science Research Institute,
Foreign Large-Size Research Facility Application Supporting project,
the Global Science Experimental Data Hub Center, the Korea Institute of
Science and Technology Information (K24L2M1C4)
and
KREONET/GLORIAD;
Universiti Malaya RU grant, Akademi Sains Malaysia, and Ministry of Education Malaysia;
Frontiers of Science Program Contracts
No.~FOINS-296,
No.~CB-221329,
No.~CB-236394,
No.~CB-254409,
and
No.~CB-180023, and SEP-CINVESTAV Research Grant No.~237 (Mexico);
the Polish Ministry of Science and Higher Education and the National Science Center;
the Ministry of Science and Higher Education of the Russian Federation
and
the HSE University Basic Research Program, Moscow;
University of Tabuk Research Grants
No.~S-0256-1438 and No.~S-0280-1439 (Saudi Arabia), and
King Saud University,Riyadh, Researchers Supporting Project number (RSPD2024R873)  
(Saudi Arabia);
Slovenian Research Agency and Research Grants
No.~J1-9124
and
No.~P1-0135;
Ikerbasque, Basque Foundation for Science,
the State Agency for Research of the Spanish Ministry of Science and Innovation through Grant No. PID2022-136510NB-C33,
Agencia Estatal de Investigacion, Spain
Grant No.~RYC2020-029875-I
and
Generalitat Valenciana, Spain
Grant No.~CIDEGENT/2018/020;
the Swiss National Science Foundation;
The Knut and Alice Wallenberg Foundation (Sweden), Contracts No.~2021.0174 and No.~2021.0299;
National Science and Technology Council,
and
Ministry of Education (Taiwan);
Thailand Center of Excellence in Physics;
TUBITAK ULAKBIM (Turkey);
National Research Foundation of Ukraine, Project No.~2020.02/0257,
and
Ministry of Education and Science of Ukraine;
the U.S. National Science Foundation and Research Grants
No.~PHY-1913789 
and
No.~PHY-2111604, 
and the U.S. Department of Energy and Research Awards
No.~DE-AC06-76RLO1830, 
No.~DE-SC0007983, 
No.~DE-SC0009824, 
No.~DE-SC0009973, 
No.~DE-SC0010007, 
No.~DE-SC0010073, 
No.~DE-SC0010118, 
No.~DE-SC0010504, 
No.~DE-SC0011784, 
No.~DE-SC0012704, 
No.~DE-SC0019230, 
No.~DE-SC0021274, 
No.~DE-SC0021616, 
No.~DE-SC0022350, 
No.~DE-SC0023470; 
and
the Vietnam Academy of Science and Technology (VAST) under Grants
No.~NVCC.05.12/22-23
and
No.~DL0000.02/24-25.

These acknowledgements are not to be interpreted as an endorsement of any statement made
by any of our institutes, funding agencies, governments, or their representatives.

We thank the SuperKEKB team for delivering high-luminosity collisions;
the KEK cryogenics group for the efficient operation of the detector solenoid magnet and IBBelle on site;
the KEK Computer Research Center for on-site computing support; the NII for SINET6 network support;
and the raw-data centers hosted by BNL, DESY, GridKa, IN2P3, INFN, 
PNNL/EMSL, 
and the University of Victoria.
\end{acknowledgments}

\end{document}


\vspace*{1\baselineskip}


\title{Supplemental  material}



\maketitle

\section{Supplemental Figures}

\begin{figure*}[b!]
\includegraphics[scale=0.35]{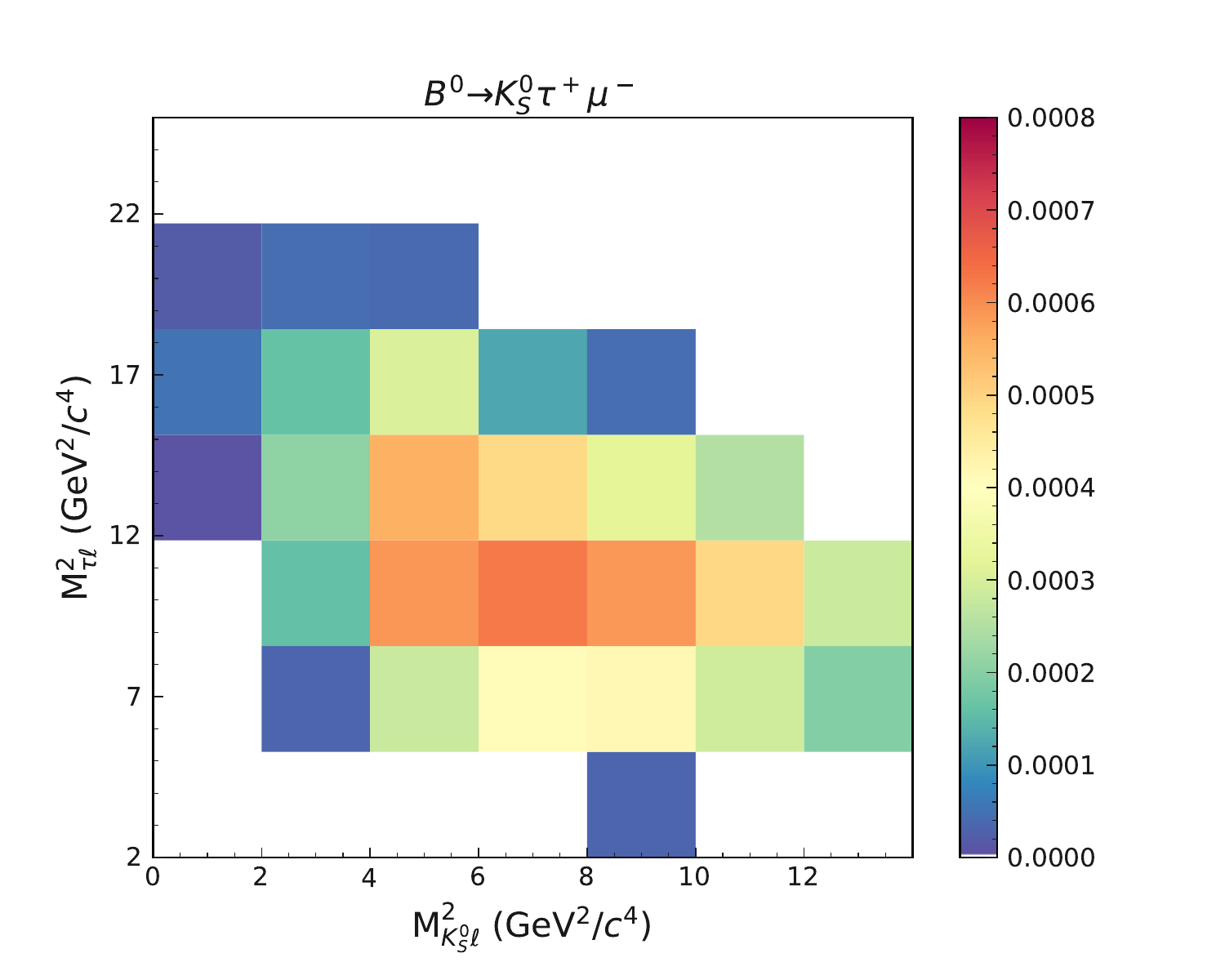}
\includegraphics[scale=0.35]{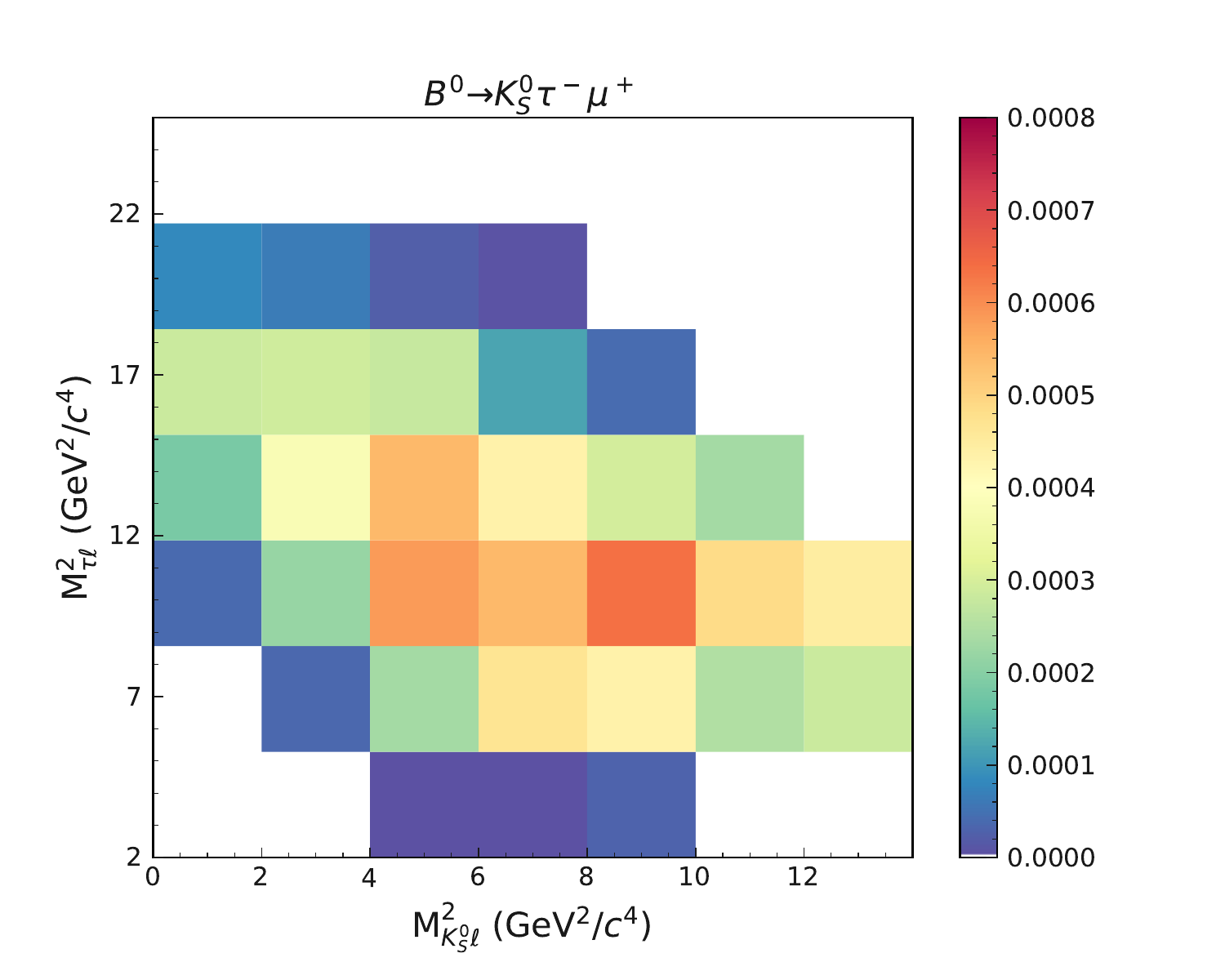}\\
\includegraphics[scale=0.35]{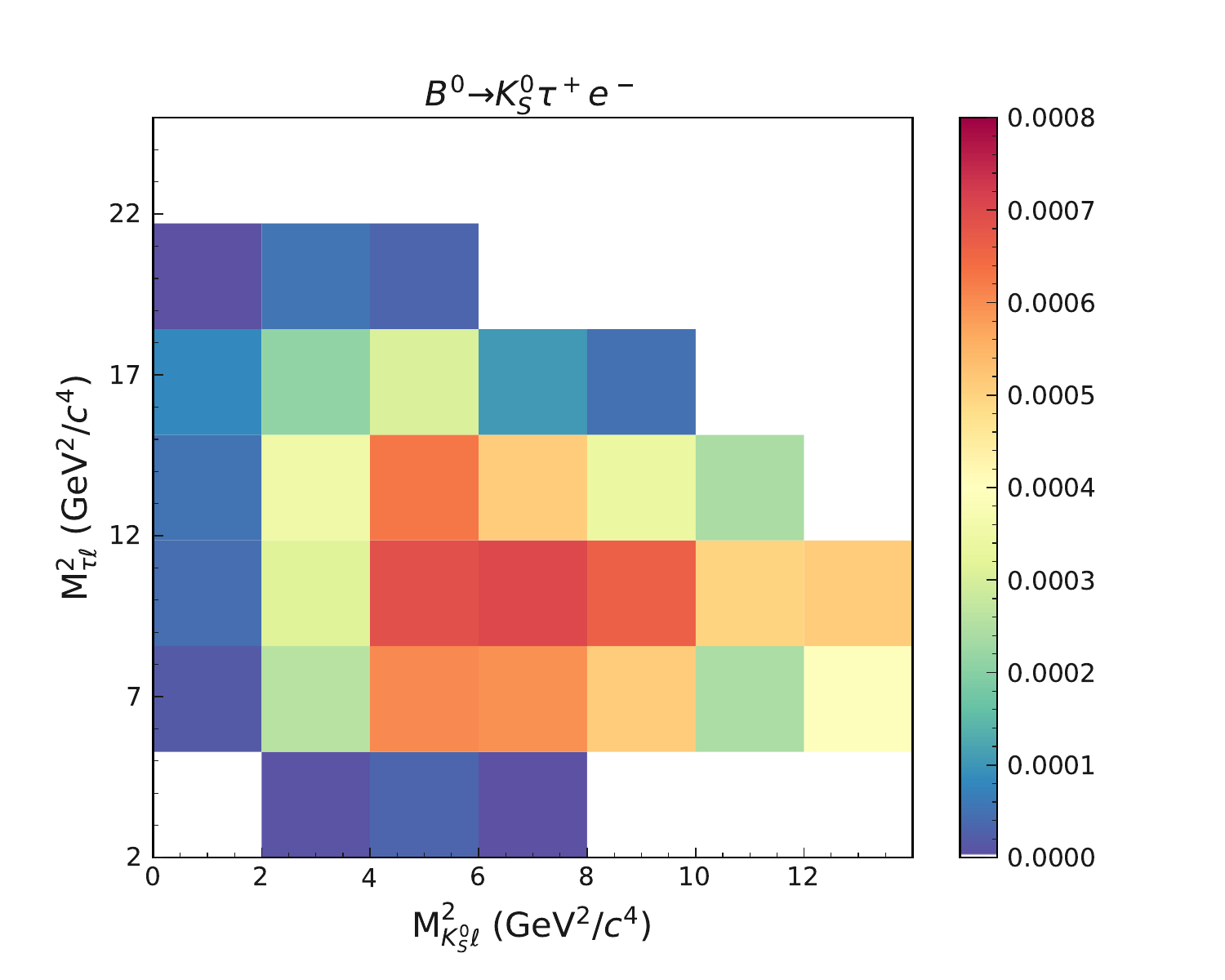}
\includegraphics[scale=0.35]{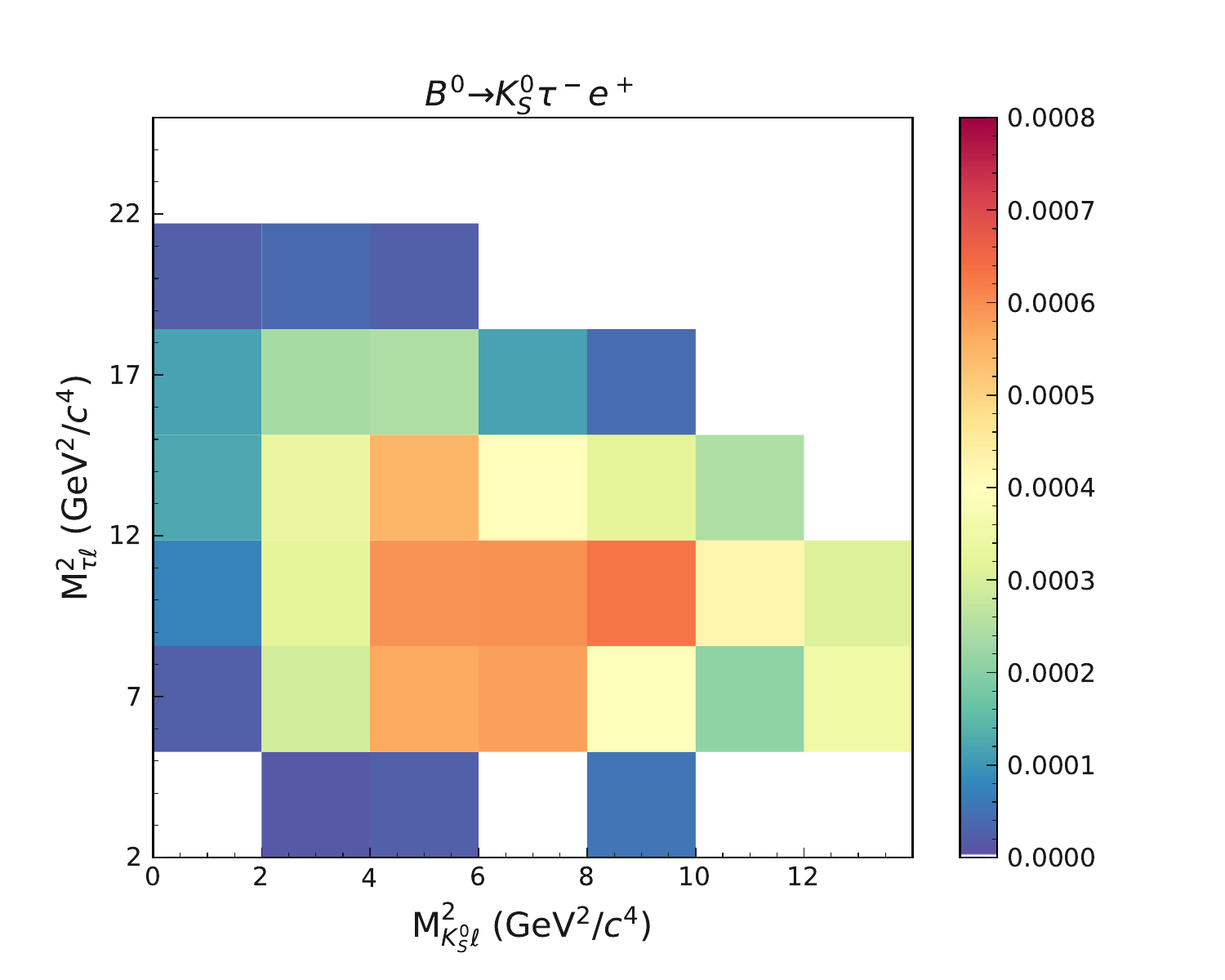}
\caption{Selection efficiency as a function of the kinematic variables $M^2 (\tau \ell)$ and $M^2 (K_S^0\ell)$. \label{fig:effMap}}
\end{figure*}

Figure~\ref{fig:effMap} presents the selection efficiencies for the four signal modes: $B^{0}\!\! \rightarrow \!K^{0}_S\tau^{+}\mu^{-}$, $B^{0}\!\! \rightarrow \!K^{0}_S\tau^{-}\mu^{+}$, $B^{0}\!\! \rightarrow \!K^{0}_S\tau^{+}e^{-}$, and $B^{0}\!\! \rightarrow \!K^{0}_S\tau^{-}e^{+}$. The efficiency is shown as a function of two kinematic variables, $M^2_{\tau \ell}$ and $M^2_{K_S^0\ell}$, where the 4-momentum of the $\tau$ lepton is inferred using the $B_{\rm tag}$ reconstruction. These distributions can be utilized to reinterpret the results for different models and kinematics, extending beyond the uniform phase space distribution assumed in the signal simulation and upper limit estimation.
